\documentclass[12pt]{article}

\parskip=\medskipamount            
\def\7#1#2{\mathop{\null#2}\limits^{#1}}        

\def\beee{\begin{equation}}
\def\eeee{\end{equation}}

\begin{document}
\begin{center}
\textbf{FAILURE OF MICROCAUSALITY IN QUANTUM FIELD THEORY ON NONCOMMUTATIVE SPACETIME}\\
[5mm]
O.W. Greenberg\footnote{email address: owgreen@physics.umd.edu.}\\
{High Energy Physics Division, Department of Physical Sciences,\\
University of Helsinki, FIN-00014, Helsinki, Finland}\footnote{permanent address
\it Center for Theoretical Physics,
Department of Physics,
University of Maryland,
College Park, MD~~20742-4111, USA}\\
~\\
\end{center}

\bibliographystyle{unsrt}

\begin{abstract}

The commutator of $:\phi(x) \star \phi(x):$ with
$\partial^y_{\mu}:\phi(y) \star \phi(y):$ fails to vanish at equal times and
thus also fails to obey microcausality at spacelike separation
even for the case in which $\theta^{0i}=0$. The failure to obey
microcausality for these sample observables implies that this
form of noncommutative field theory fails to obey microcausality
in general. This result holds generally when there are time derivatives
in the observables. We discuss possible responses to this problem.

\end{abstract}

\section{Introduction}
   There is broad agreement that new possibilities, beyond the
standard model, must be explored to understand how to reconcile
relativistic quantum theory with the theory
of gravity provided by Einstein's general theory of relativity,
as well as to reduce the number of parameters that
must be found empirically in order to make the standard model
precise. String theory is the most far-reaching of the extensions
of the standard model. Quantum field theory on noncommutative spacetime
stands as an intermediate framework between string theory and the
usual quantum theory of fields. Noncommutative spacetime was considered
as long ago as 1947~\cite{sny}. Because this intermediate theory is
more manageable than string theory, quantum field theory on noncommutative
spacetime has aroused a good deal of interest following the work by
Doplicher, et al~\cite{dop}. Reviews appear in~\cite{dou, sza}. The specific type of
theory on noncommutative spacetime that has been studied the most is
the one in which the noncommutativity takes the form
\beee
[\hat{x}^{\mu}, \hat{x}^{\nu}]_-=i\theta^{\mu \nu}         \label{cr}
\eeee
with $\theta^{\mu \nu}$ chosen to be a constant matrix. Most authors
make this choice only for the case Eq.(\ref{cr}). Some authors also
assume
\beee
[\hat{x}^{\mu}, \hat{y}^{\nu}]_-=i\theta^{\mu \nu}, y \neq x.  \label{xy}
\eeee
The argument for Eq.(\ref{cr}) is that it follows both from string
theory in a background ``magnetic'' field~\cite{wit,sei} 
and from the equation for the
motion of an electron in a magnetic field~\cite{lan}. The argument that
one should also adopt Eq.(\ref{xy}) because otherwise there would be
a discontinuity for $\hat{x} \rightarrow \hat{y}$ seems to be based on
too naive an interpretation of the symbol $\hat{x}$. We will not adopt
Eq.(\ref{xy}); rather we will assume
\beee
[\hat{x}^{\mu}, \hat{y}^{\nu}]_-=0, y \neq x  \label{nxy}
\eeee
for most of our discussion. Since some authors do use the star product
that follows for $x \neq y$ both for field products and in between the
terms of a commutator~\cite{asc,dim,cnt}, we will discuss this case in a later section.
With the assumption of Eq.(\ref{cr}) we replace the field 
$\phi(\hat{x})$ by $\phi(x)$ and use the star product~\cite{dou,sza} for
the product of fields at the {\em same} spacetime point.
This means that field theory
on noncommutative spacetime becomes a particular
form of nonlocal field theory, with the nonlocality expressed in 
terms of the Moyal
phases that occur in the star product. 

One of the major problems with this case of constant $\theta^{\mu \nu}$
is that it breaks the Lorentz group $SO(1,3)$ to $SO(1,1) \times SO(2)$ which
is abelian and thus has only one-dimensional irreducible representations.
Because of this, no spinor, vector, etc. fields would exist.  
M. Chaichian, et al~\cite{cha, chapre}, 
J. Wess~\cite{wes} and P. Aschieri, et al~\cite{asc}, have shown that the theory has a 
twisted Lorentz
(and also Poincar\'{e}) symmetry in which the full $SO(1,3)$ symmetry 
remains, and
thus the spinor, vector, etc. representations do occur. To date the full
significance of this twisted symmetry is unclear.

In this paper we consider the question of microcausality of observables; i.e.,
of vanishing of the commutator of observables at spacelike separation. This
condition is often called locality, but since locality can have several meanings,
we will use ``microcausality" for this requirement. 

Chaichian, et al~\cite{chanis}, studied microcausality for the choice of
${\mathcal O}(x) \equiv :\phi(x) \star \phi(x):$ as a sample observable
and found that it obeys microcausality provided that $\theta^{0i}=0$.
We will take $\theta^{0i}=0$ throughout this paper~\cite{car}.
Since this condition is required for unitarity, this is not a further
restriction on the theory. These authors expected that microcausality
would hold generally for observables, but we show below that this is not
the case.
Because microcausality should hold for all observables, we also want
$\partial_{\mu}{\mathcal O}(x)$ to obey microcausality relative to $\mathcal{O}(y)$
as well as relative to $\partial_{\nu}\mathcal{O}(y)$. We find that microcausality
fails for some of these cases.

In the discussion of microcausality, to prove a positive result
one must show that all matrix elements of the commutator obey microcausality. To show
a negative result, that the commutator violates microcausality, one need only show
that any single matrix element of the commutator violates microcausality.

Although we give detailed calculations for the sample observable considered by
Chaichian, et al, our results are valid for any fields or observables, as we discuss
later.

\section{Calculation of $[{\mathcal O}(x), \partial_{\nu}{\mathcal O}(y)]_-$}
Here are our normalization and other conventions which differ from
those of Chaichian, et al,\\
\beee
\phi(x)=(2 \pi)^{-D/2} \int \tilde{\phi}(k) e^{-ik \cdot x}d^D k,
\eeee
\beee
\langle0|\tilde{\phi}(k) \tilde{\phi}(l)|0\rangle=
2 \pi \theta(k^0) \delta(k^2-m^2) \delta(k+l),
\eeee
\beee
\langle0|\tilde{\phi}(k)|p\rangle=\delta(k-p),~~~E_k=\sqrt{\mathbf{k}^2+m^2}.
\eeee
We find
\begin{eqnarray}
\lefteqn{
\langle 0|[:\phi(x) \star \phi(x):, :\phi(y) \star \phi(y):]_-|p,p^{\prime}\rangle=}
      \nonumber \\
& &
(e^{-i p \cdot x -i p^{\prime} \cdot y} + e^{-i p^{\prime} \cdot x -i p \cdot y})
\frac{4}{(2\pi)^{2D-1}} \int d^D k
\epsilon(k^0)\delta(k^2-m^2)e^{-i k \cdot (x-y)}\times   \nonumber   \\
& & \cos (\frac{1}{2} \theta^{\mu \nu}k_{\mu}
p_{\nu}) \cos (\frac{1}{2} \theta^{\mu \nu}k_{\mu} p_{\nu}^{\prime}),  \label{chanis}
\end{eqnarray}
which agrees, up to irrelevant 
numerical factors, with the calculation of Chaichian, et al~\cite{chanis}.
We also calculated the anticommutator, obtained by replacing
$\epsilon(k^0)$ by $1$ in Eq.(\ref{chanis}), and also checked by direct calculation
\begin{eqnarray}
\lefteqn{\langle 0|[:\phi(x) \star \phi(x):, :\phi(y) \star \phi(y):]_+|p,p^{\prime}\rangle=}
      \nonumber \\
& &
(e^{-i p \cdot x -i p^{\prime} \cdot y} + e^{-i p^{\prime} \cdot x -i p \cdot y})
\frac{4}{(2\pi)^{2D-1}} \int d^D k
\delta(k^2-m^2)e^{-i k \cdot (x-y)}\times   \nonumber   \\
& & \cos (\frac{1}{2} \theta^{\mu \nu}k_{\mu}
p_{\nu}) \cos (\frac{1}{2} \theta^{\mu \nu}k_{\mu} p_{\nu}^{\prime}).
\end{eqnarray}
From Eq.(\ref{chanis}),
\begin{eqnarray}
\lefteqn{[{\mathcal O}(x), \partial_{\nu}{\mathcal O}(y)]_- = }  \nonumber  \\
& &
-i(p^{\prime}_{\nu}e^{-i p \cdot x -i p^{\prime} \cdot y} + 
p_{\nu}e^{-i p^{\prime} \cdot x -i p \cdot y})
\frac{4}{(2\pi)^{2D-1}} \int d^D k
\epsilon(k^0)\delta(k^2-m^2)e^{-i k \cdot (x-y)}\times   \nonumber   \\
& &  \cos (\frac{1}{2} \theta^{\mu \nu}k_{\mu}p_{\nu}) 
\cos (\frac{1}{2} \theta^{\mu \nu}k_{\mu} p_{\nu}^{\prime}) + \nonumber\\
& &
 (e^{-i p \cdot x -i p^{\prime} \cdot y} + e^{-i p^{\prime} \cdot x -i p \cdot y})
\frac{4}{(2\pi)^{2D-1}} \int d^D k (i k_{\nu})
\epsilon(k^0)\delta(k^2-m^2)e^{-i k \cdot (x-y)}\times   \nonumber   \\
& &
 \cos (\frac{1}{2} \theta^{\mu \nu}k_{\mu}p_{\nu}) 
\cos (\frac{1}{2} \theta^{\mu \nu}k_{\mu} p_{\nu}^{\prime}).  
\end{eqnarray}
At $x^o=y^0$, the $\nu=0$ term is
\beee
(e^{-i p \cdot x -i p^{\prime} \cdot y} + e^{-i p^{\prime} \cdot x -i p \cdot y})
\frac{4i}{(2\pi)^{2D-1}} \int d^{D-1} k 
e^{i\mathbf{k} \cdot (\mathbf{x}-\mathbf{y})}
\cos (\frac{1}{2} \theta^{ij}k_{i}
p_{j}) \cos (\frac{1}{2} \theta^{ij}k_{i} p_{j}^{\prime}).   \label{dcrnstar}
\eeee
In order for this to vanish for $\mathbf{x}-\mathbf{y} \neq 0$, the Fourier transform of
Eq.(\ref{dcrnstar}) must be a {\em polynomial} in $k$. Since this Fourier transform
is $\cos (\frac{1}{2} \theta^{ij}k_{i}p_{j}) 
\cos (\frac{1}{2} \theta^{ij}k_{i} p_{j}^{\prime})$, it is not a polynomial in
$k$, and thus this commutator violates microcausality. If we carry out the $\int d^{D-1}k$ we
get a sum of delta functions that exhibits the violation of microcausality explicitly. 
To be explicit, let $\theta^{12}=-\theta^{12}=\theta$, other values of $\theta=0$, then,
up to irrelevant factors, the nonlocality is
\beee
\sum_{s=\pm 1, t=\pm 1}\delta(x^1-y^1-s \theta (p_2+t p_2^{\prime})) 
\delta(x^2-y^2-s\theta (p_1+t p_1^{\prime}))\delta(x^3-y^3).  \label{grows}
\eeee
The nonlocality increases with the sum or difference of the momenta of the particles.

We expect that the commutator of any observable which is a polynomial in free fields with
odd numbers of time derivatives will fail to commute at spacelike separation, 
just as in the case
calculated above. A relevant case of such an observable is the current of a charged scalar field.
The basic reason for these violations of spacelike commutativity is that the space averaging of 
zero, which occurs for $\Delta(x-y)$ at $x^0=y^0$, is still zero. By contrast, the space
averaging of $\delta(\mathbf{x}-\mathbf{y})$, which occurs for 
$\partial_{x^0}\Delta(x-y)$ at $x^0=y^0$, is not zero.
 
\section{Calculation of a matrix element of the star commutator}

In the study of $[{\mathcal O}(x), {\mathcal O}(y)]_-$ Chaichian, et al~\cite{chanis} considered
the ordinary commutator rather than the star commutator,
\begin{eqnarray}
\lefteqn{[{\mathcal O}(x), {\mathcal O}(y)]_{\star -} = 
[:\phi(x) \star \phi(x):, :\phi(y) \star \phi(y):]_{\star -} \equiv} \nonumber \\
 & &
:\phi(x) \star \phi(x): \star :\phi(y) \star \phi(y):-:\phi(y) \star \phi(y):
\star :\phi(x) \star \phi(x):.
\end{eqnarray}
We have calculated the star commutator for this sample observable. We anticipate
that the star commutator will give a qualitatively different result than
the ordinary one, because the Moyal phases in the star commutator
will be sensitive to both coordinates
$x$ and $y$ and thus to the separation of $x$ and $y$, while the star product
in the observable itself is not aware of this separation. From a more
calculational point of view, the new Moyal phases in the terms of the star
commutator will have opposite sign in the two terms. Thus if the Moyal phase
in one term of the star commutator is $e^{i \Theta}$ the phase in the
other term will be $e^{-i \Theta}$ and the star commutator will have the
form
\beee
[{\mathcal O}(x), {\mathcal O}(y)]_{\star -} 
=\cos \Theta [{\mathcal O}(x), {\mathcal O}(y)]_ - +i \sin \Theta
[{\mathcal O}(x), {\mathcal O}(y)]_+,     \label{main}
\eeee
where $\Theta$ is the differential operator,
\beee
\Theta = \frac{i}{2} \theta^{\mu \nu} \partial_{\mu}^x \partial_{\nu}^y.  \label{Theta}
\eeee
The anticommutator term will not vanish at spacelike separation.
We only have to convert the differential operator, $\Theta$, defined
in Eq.(\ref{Theta}), to momentum space and insert it in Eq.(\ref{main}) to find
\begin{eqnarray}
\lefteqn{[{\mathcal O}(x), {\mathcal O}(y)]_{\star -} = }  \nonumber  \\
 & &
(e^{-i p \cdot x -i p^{\prime} \cdot y} + e^{-i p^{\prime} \cdot x -i p \cdot y}) \times \\
 & & \frac{4}{(2 \pi)^{2D-1}}
\int d^D k (\epsilon(k) \cos \tilde{\Theta} +i \sin \tilde{\Theta)} \delta(k^2-m^2)e^{-i k \cdot (x-y)}
\times  \nonumber  \\
 & & \cos (\frac{1}{2} \theta^{\mu \nu}k_{\mu}
p_{\nu}) \cos (\frac{1}{2} \theta^{\mu \nu}k_{\mu} p_{\nu}^{\prime}),
\end{eqnarray}
where now
\beee
\tilde{\Theta} = -\frac{1}{2} \theta^{ij}(k_i (p+p^{\prime})_j + p_i p^{\prime}_j).  \label{ptheta}
\eeee
At equal times, 
\begin{eqnarray}
\lefteqn{[{\mathcal O}(x), {\mathcal O}(y)]_{\star - ET} = }  \nonumber  \\
 & &
(e^{-i p \cdot x -i p^{\prime} \cdot y} + e^{-i p^{\prime} \cdot x -i p \cdot y}) \times \\
 & & \frac{4}{(2 \pi)^{2D-1}}
\int \frac{d^{D-1} k}{2E_k}[ \cos \tilde{\Theta}(e^{i\mathbf{k} \cdot (\mathbf{x}-\mathbf{y})}
-e^{i\mathbf{k} \cdot (\mathbf{x}-\mathbf{y})}) +
i \sin \tilde{\Theta} (e^{i\mathbf{k} \cdot (\mathbf{x}-\mathbf{y})}+
e^{i\mathbf{k} \cdot (\mathbf{x}-\mathbf{y})})]
\times  \nonumber  \\
 & & \cos (\frac{1}{2} \theta^{ij}k_{i}p_{j}) \cos (\frac{1}{2} \theta^{ij}k_{i} p_{j}^{\prime}),
\end{eqnarray}
where we have exhibited explicitly the contributions from the two mass shells. Obviously
the coefficient of the $\cos \tilde{\Theta}$ term vanishes. The final result is
\begin{eqnarray}
\lefteqn{[{\mathcal O}(x), {\mathcal O}(y)]_{\star - ET} = }  \nonumber  \\
 & &
(e^{-i p \cdot x -i p^{\prime} \cdot y} + e^{-i p^{\prime} \cdot x -i p \cdot y}) \times \\
 & & \frac{8i}{(2 \pi)^{2D-1}}
\int \frac{d^{D-1} k}{2E_k}
 \sin (-\frac{1}{2}\theta^{ij}(k_i(p+p^{\prime})_j+p_ip^{\prime}_j)
 (e^{i\mathbf{k} \cdot (\mathbf{x}-\mathbf{y})})
\times  \nonumber  \\
 & & \cos (\frac{1}{2} \theta^{ij}k_{i}p_{j}) \cos (\frac{1}{2} \theta^{ij}k_{i} p_{j}^{\prime}).
\end{eqnarray}
As in the previous section, the Fourier transform of this is not a polynomial in $k$,
so this quantity does not vanish at spacelike separation.  
Clearly if we drop the $\tilde{\Theta}$
term we recover the result for the ordinary commutator which does obey microcausality. (We can
equip the $\tilde{\Theta}$ parameter with a factor $\lambda$ if we want to go continuously 
between the ordinary and the star commutator.) This completes the demonstration
that the star commutator of this sample observable does not vanish at spacelike
separation.

\section{Star commutator and anticommutator of general fields and  observables}
The analog of the result we gave in Eq.(\ref{main}) holds for any fields and observables. 
Although our discussion above concerned neutral scalar fields, our conclusions hold for
fields, neutral or charged, of any spin provided the usual connection of spin and type of
commutation relation (using a commutator or an anticommutator) is used.
For fields or observables whose commutators vanish at spacelike separation in ordinary
field theory, the star commutators of the fields or observables on noncommutative
spacetime fail to vanish. Correspondingly, for fields or observables whose 
anticommutators vanish at spacelike separation in ordinary
field theory, the star anticommutators of the fields or observables on noncommutative
spacetime fail to vanish. 
Thus even the free
field star commutator (anticommutator) does not vanish at spacelike separation. 
Because of the anticommutator (commutator)
term in Eq.(\ref{main}) the free field commutator (anticommutator) 
on noncommutative spacetime is neither translation invariant
nor a c-number. On the other hand, the 
vacuum matrix element, and thus also the propagator 
of the free field on noncommutative spacetime, is the usual one, because
for the vacuum matrix element the derivatives in Eq.(\ref{Theta}) or the momenta in 
Eq.(\ref{ptheta}) are linearly dependent so that the Moyal phase vanishes.

\section{Related work}
The increase of nonlocality with momentum that we found in Eq.(\ref{grows}) is 
similar to that found by~\cite{seisuss}. 
H. Bozkaya, et al,~\cite{boz} studied microcausality in noncommutative field theory in
the context of perturbation theory using different definitions of time-ordering and
concluded that microcausality and unitarity are in conflict. Our simpler calculation was
done in the context of free field observables rather than in perturbation theory.
L. Alvarez-Gaume', et al,~\cite{alv} also studied microcausality in perturbation
theory and found that $SO(1,3)$ microcausality is violated but that $SO(1,1)$ microcausality,
i.e. microcausality in the light wedge,
holds if and only if perturbative unitarity holds.

\section{Comments about the failure of microcausality}
Since the light cone has no status in a theory with constant $\theta^{\mu \nu}$
it is surprizing that microcausality
can hold in some special cases, such as the case in which the observables are
constructed from scalar fields with no time derivatives~\cite{chanis}. What one should
expect is that only the light wedge, $x^{0~2}-x^{3~2} \leq 0$, has
meaning as discussed by~\cite{alv}. Assuming $\theta^{12} = \theta$, 
with the other elements of the $\theta$ matrix
equal to zero, both the ordinary and the star commutator of observables vanish
trivially in the light wedge. Very likely the choice of constant $\theta$ should be
abandoned in favor of a $\theta$ that transforms under the Lorentz group. This was the
point of view of Snyder~\cite{sny} in his early work and recently 
has been suggested by Doplicher, et al~\cite{dop,dop2}.
Other responses to the failure of microcausality that we
demonstrated in the previous sections include: (a) that massive
string states cannot be neglected in quantum field theory on noncommutative spacetime,
at least in the version in which the noncommutativity occurs as a constant matrix
$\theta^{\mu \nu}$ as in Eq.(\ref{cr}), and the noncommutativity is implemented via
the star product as described above. Gomis and Mehen~\cite{gom} have shown that theories
with electric ($\theta^{0i} \neq 0$) noncommutativity violate unitarity, except for
the case of lightlike noncommutativity~\cite{aha}, and do
not represent a low-energy limit of string theory, while theories, at least in
perturbation theory to one loop, with magnetic ($\theta^{0i} = 0, \theta^{ij} \neq 0$)
noncommutativity
obey unitarity and can serve as a low-energy limit of string theory. The situation
here differs from the case considered by Gomis and Mehen, not only because microcausality
is at stake instead of unitarity, but also because the problem occurs even when
$\theta^{0i} = 0$. Nonetheless, the results of Gomis and Mehen may give a hint that
the problem arises because of the neglect of massive string states. If this is the
correct way to understand the failure of microcausality, we should ask if there is
some way to amend the usual space-space noncommutativity so that the massive string
modes can be incorporated and microcausality can be restored; or (b) to drop the
requirement of microcausality. We do not speculate on the implications of this last
response in this paper.


Acknowledgements: We are happy to thank Masud Chaichian for stimulating
our interest in field theory on noncommutative spacetime, for many helpful
discussions, and for his hospitality at the University of Helsinki. We thank
Lew Licht, 
Claus Montonen, Kazuhiko Nishijima, Anca Tureanu, Peter Pre\v{s}najder and 
Ram Sriharsha for helpful discussions, as well as an anonymous referee 
for comments that significantly improved this paper.

This work was supported in part by the National Science Foundation,
Grant No. PHY-0140301.

\end{document}